\documentclass[%
reprint,
superscriptaddress,
 amsmath,amssymb,
 aps,
prb,
]{revtex4-2}
\usepackage{amsfonts}
\usepackage{amstext}
\usepackage{amssymb}
\usepackage{amsmath}
\usepackage{bm}% bold math
\usepackage[utf8]{inputenc}
\usepackage[T1]{fontenc}
\usepackage{graphicx}
\usepackage{color,xcolor}

\usepackage{hyperref}
\usepackage{orcidlink}
\hypersetup{colorlinks=true,citecolor=blue,linkcolor=blue,filecolor=blue,urlcolor=blue,pdfstartview=FitH,pdfpagemode=UseNone}
\usepackage{todonotes}
\usepackage[normalem]{ulem}
\usepackage{soul}
\newcommand{\be}{\begin{equation}}
\newcommand{\ee}{  \end{equation}}
\newcommand{\ba}{\begin{eqnarray}}
\newcommand{\ea}{  \end{eqnarray}}
\newcommand{\ve}{\varepsilon}

\bibpunct{[}{]}{,}{n}{}{}
\begin{document}

\title{
%NESS Entropy Production in the Strong Coupling Regine\\or\\
Entropy Production in Mesoscopic Electronic Transport Beyond Weak-Coupling}

\author{Ivana R. A. C. Lucena\,\orcidlink{0000-0002-3205-4228}}
\affiliation{Instituto de F\'isica Te\'orica,
Universidade Estadual Paulista (UNESP),
S\~ao Paulo, SP, Brazil}

\author{Caio H. Lewenkopf\,\orcidlink{0000-0002-2053-2798}}
\affiliation{Instituto de F\'isica,
Universidade Federal do Rio de Janeiro,
21941-972 Rio de Janeiro, RJ, Brazil}
\affiliation{Institut für Theoretische Physik,
Universität Regensburg,
D-93040 Regensburg, Germany}

\author{Alexandre R. Rocha\,\orcidlink{0000-0001-8874-6947}}
\affiliation{Instituto de F\'isica Te\'orica,
Universidade Estadual Paulista (UNESP),
S\~ao Paulo, SP, Brazil}

\date{\today}

\begin{abstract}
We investigate the full statistics of entropy production in coherent mesoscopic conductors beyond the weak-coupling regime. Using the inside-outside perspective and the Landauer-B\"uttiker scattering framework, we relate entropy production to the full counting statistics of particle transport and show that the corresponding cumulant-generating function satisfies the fluctuation symmetry independently of the coupling strength. We obtain closed analytical expressions for the cumulant-generating function in the weak- and strong-coupling limits, enabling a detailed characterization of entropy production distributions and higher-order cumulants. These cumulants reveal distinct signatures of resonant transport and coherent fermionic partitioning. While coupling-induced broadening substantially modifies the transport statistics, the fluctuation symmetry imposed by microscopic reversibility remains preserved in the long-time limit.
\end{abstract}

\maketitle

%%%%%%%%%%%%%%%%%%%%%%%%%%%%%%%%%%%%%%%%%%%%%%%%%%
\section{Introduction}
\label{sec:intro}
%%%%%%%%%%%%%%%%%%%%%%%%%%%%%%%%%%%%%%%%%%%%%%%%%%

In mesoscopic conductors, quantum transport is typically characterized by charge, energy, and heat currents. Beyond these primary observables, entropy production $\sigma$ plays a distinguished role as a direct measure of irreversibility~\cite{Landi2021}. Its statistical properties have become a cornerstone of quantum thermodynamics, revealing fundamental connections between nonequilibrium transport, thermodynamics, and information theory~\cite{Seifert2012,Benenti2017,Santos2019,Esposito2010,Strasberg2017PRX,Guarnieri2019,Pekola2021}. 
Beyond fundamental considerations, entropy production and its fluctuations govern the performance bounds of nanoscale thermal devices, including thermoelectric heat engines and refrigerators~\cite{Sivan1986,Sanchez2011}. 
Remarkable experimental progress, such as the direct measurement of entropy production and local dissipation in mesoscopic systems~\cite{wang2002experimental, liphardt2001equilibrium, Hartman2018, Josefsson2018, Marguerite2019, Pyurbeeva2021, Pekola2021, Gehring2021, Child2022}, makes these theoretical advances increasingly accessible to observation.

A landmark result in this context is the fluctuation theorem (FT), which establishes a universal symmetry for the probability distribution of entropy production arbitrarily far from equilibrium~\cite{evans1993probability, gallavotti1995dynamical, evans1994equilibrium, lebowitz1999gallavotti, jarzynski1997nonequilibrium, crooks1999entropy, van2004power, spinney2012fluctuation}. 
By quantifying the relative probability of positive and negative entropy production events, the FT has reshaped our understanding of irreversible processes and provided the conceptual foundation for several developments, such as thermodynamic uncertainty relations and related kinetic bounds~\cite{Horowitz2017,di2019kinetic,hasegawa2020quantum,prech2025role}. 

Despite these advances, most formulations of quantum thermodynamics rely on the assumption of weak system--environment coupling, where thermodynamic quantities are well defined and the distinction between system and reservoir remains unambiguous. 
Extending this framework to intermediate and strong coupling regimes, however, remains a major challenge ~\cite{Esposito2015a,Bruch2016}. 
This issue is particularly relevant for coherent mesoscopic conductors, where transport routinely occurs beyond the weak-coupling limit.

%------------------------- FIGURE 1 ---------------------------------
\begin{figure}[ht]
    \centering

\includegraphics[width=0.9\columnwidth]{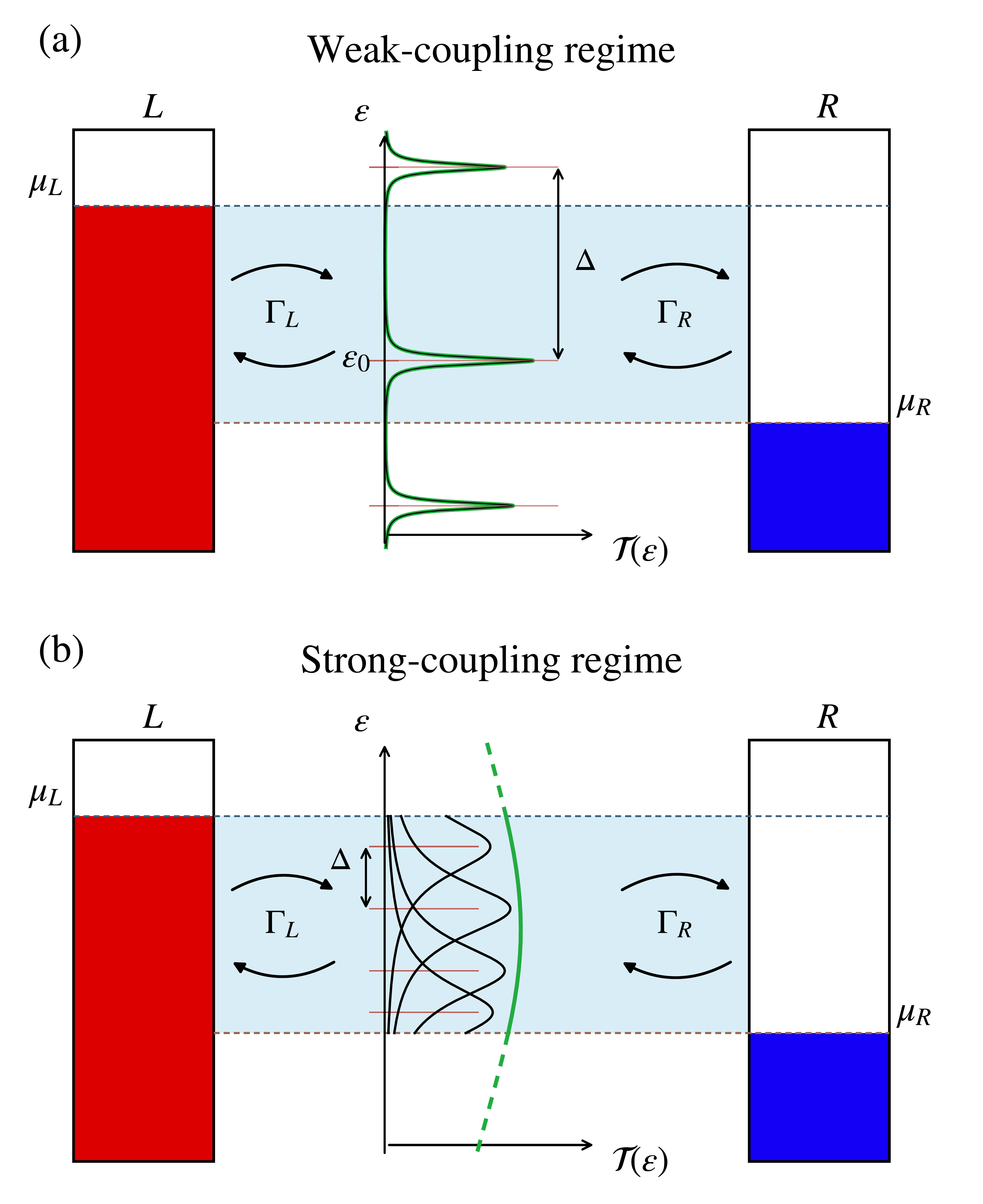}
\caption{Schematic representation of a two-terminal mesoscopic conductor connected to left ($L$) and right ($R$) electronic reservoirs. 
The solid green curve illustrates the total energy-dependent transmission $\mathcal{T}(\varepsilon)$, where $\mu_{\alpha}$ denotes the chemical potential and $\Gamma_{\alpha}$ the coupling to lead $\alpha \in \{L, R\}$. 
Thin black lines depict the line-shape profiles of individual resonant levels with average level spacing $\Delta$. 
(a) Weak-coupling regime ($\Gamma \ll \Delta$), characterized by well-isolated single-level resonances. 
(b) Strong-coupling regime ($\Gamma \gtrsim \Delta$), featuring strongly overlapping resonances with coherent level broadening.}

    \label{fig:setup}
\end{figure}

An important advance in this direction was made by Ref.~\cite{Bruch2018} where quantum thermodynamics for nonequilibrium steady states was formulated within an inside-outside perspective using the Landauer--B\"uttiker scattering approach \cite{Imry1999, Datta1995, Buttiker1986, Blanter2000,Kobayashi2021} and its generalizations \cite{Galperin2024, tesser2026}. 
%\Iv{\cite{PhysRevB.46.12485,Blanter2000,Landauer01041970,Buttiker1985}}. 
In this framework, entropy production is expressed in terms of asymptotic particle and energy currents flowing through the reservoirs, providing a thermodynamically consistent description even beyond the weak-coupling regime. 
This result is fully consistent with the information-theoretic formulation~\cite{Ptaszynski2019}, in which irreversibility stems from the accumulation of correlations between the system and its environment.

While the mean entropy production beyond weak coupling can be consistently defined in terms of asymptotic currents (measured deep inside the contacts), much less is known about its fluctuations. 
In particular, an important open question is whether coherent strong coupling merely reshapes the entropy production statistics or can also affect the fluctuation symmetry underlying the FT. 
Here, we address this question within the Landauer–B\"uttiker scattering framework. 

Building on the full counting statistics of particle transport~\cite{Levitov1993, Levitov1996, Nazarov2003,
Bagrets2003, Braggio2006, Nazarov2009}, we map particle transfer fluctuations onto entropy production statistics and demonstrate that the corresponding cumulant-generating function satisfies the fluctuation symmetry independently of the coupling regime.
We then obtain analytical expressions for the full counting statistics in the weak- and strong-coupling limits and use them to characterize the corresponding entropy production distributions and higher-order cumulants. 
Our results reveal that, while coherent broadening substantially modifies the profile and dynamics of the entropy production distribution, the fluctuation theorem remains strictly preserved in the steady-state, long-time limit.

The remainder of this paper is organized as follows. 
In Sec.~\ref{sec:background} we briefly review the key literature results relevant to our study. 
In Sec.~\ref{sec:results} we present and discuss our main findings for the entropy production distribution in both the weak- and strong-coupling regimes. 
Finally, in Sec.~\ref{sec:conclusions}, we summarize our main results and offer concluding remarks.

%%%%%%%%%%%%%%%%%%%%%%%%%%%%%%%%%%%%%%%%%%%%%%%%%%%%%%%%%%%%%%%%%%
\section{Theoretical Background}
\label{sec:background}
%%%%%%%%%%%%%%%%%%%%%%%%%%%%%%%%%%%%%%%%%%%%%%%%%%%%%%%%%%%%%%%%%%

For the isothermal two-terminal conductor considered here, illustrated in Fig.~\ref{fig:setup}, entropy production is directly related to the net particle transfer between the reservoirs~\cite{PhysRevB.75.155316}. 
This relation allows the full statistics of entropy production to be obtained from the full counting statistics of particle transport~\cite{PhysRevB.75.155316,Esposito2009,PhysRevB.78.115429}. 
We therefore begin by introducing the particle transfer statistics and subsequently establish their connection to entropy production and the fluctuation theorem~\cite{Esposito2009}. 

Let $P_{\Delta t}(N)$ denote the probability that a net number $N$ of particles is transferred from the left to the right reservoir during a measurement interval $\Delta t$. Here, $N$ is an integer-valued stochastic variable, with $N>0$ denoting particle transfer from $L$ to $R$ along the chemical potential bias ($\mu_L>\mu_R$), and $N<0$ corresponding to particle transfer against it. The full statistics of $N$ are encoded in the characteristic function~\cite{Levitov1996,Nazarov2003}
\begin{equation}
\Lambda_{\Delta t}(\chi)
=
\left\langle e^{i\chi N}\right\rangle
=
\sum_{N=-\infty}^{\infty}
P_{\Delta t}(N)e^{i\chi N},
\label{eq:characteristic_function}
\end{equation}
where $\chi$ is the counting field. 
The logarithm of the characteristic function, 
\begin{equation}
\mathcal{F}_{\Delta t}(\chi)
\equiv
\ln\Lambda_{\Delta t}(\chi),
\label{eq:CGF_definition}
\end{equation}
defines the cumulant generating function (CGF). The cumulants of the transferred particle number are obtained from successive derivatives of the CGF, 
\begin{equation}
\left\langle\!\left\langle N^n\right\rangle\!\right\rangle
=
\left.
\frac{\partial^n{\mathcal{F}}_{\Delta t}(\chi)}
     {\partial(i\chi)^n}
\right|_{\chi=0}.
\label{eq:particle_cumulants}
\end{equation}
Here, $\langle\langle\cdots\rangle\rangle$ denotes cumulants, while $\langle\cdots\rangle$ denotes ordinary moments. 
In particular, the first cumulant coincides with the mean transferred particle number, $\langle\langle N\rangle\rangle=\langle N\rangle$, whereas the second cumulant is the variance, $\langle\langle N^2\rangle\rangle=\langle N^2\rangle-\langle N\rangle^2$. 
Higher-order cumulants characterize non-Gaussian features of the transfer statistics and vanish identically for a Gaussian distribution. 

In the stationary long-time limit, the CGF becomes extensive in the measurement time, $\mathcal{F}_{\Delta t}(\chi) \simeq \Delta t\,F(\chi)$, where 
\begin{equation} F(\chi) = \lim_{\Delta t\rightarrow\infty} \frac{\mathcal{F}_{\Delta t}(\chi)}{\Delta t}
\label{eq:SCGF} 
\end{equation}
is the scaled cumulant generating function (SCGF)~\cite{TOUCHETTE20091}. 
Correspondingly, the scaled cumulants are defined as
\begin{equation} C_n \equiv \lim_{\Delta t\rightarrow\infty} \frac{ \left\langle\!\left\langle N^n\right\rangle\!\right\rangle }{\Delta t} = \left. \frac{\partial^n F(\chi)} {\partial(i\chi)^n} \right|_{\chi=0}. \label{eq:scaled_cumulants} \end{equation}
These quantities characterize the steady-state particle statistics per unit time. In terms of the central moments of $N$, the first four scaled cumulants are 
\begin{align}
C_1 &= \lim_{\Delta t\rightarrow\infty}\Delta t^{-1} \langle N\rangle = I^{N}, \label{eq:C1} \\ 
C_2 &= \lim_{\Delta t\rightarrow\infty}\Delta t^{-1} \left\langle\Delta N^2 \right\rangle, \label{eq:C2} \\ 
C_3 &= \lim_{\Delta t\rightarrow\infty} \Delta t^{-1}\left\langle \Delta N^3 \right\rangle, \label{eq:C3} \\
C_4 &= \lim_{\Delta t\rightarrow\infty} \Delta t^{-1} \left[ \left\langle \Delta N^4 \right\rangle - 3 \left\langle \Delta N^2 \right\rangle^2 \right]. \label{eq:C4} 
\end{align}
Here, $\Delta N = N - \langle N\rangle$. Thus, $C_1$ gives the average particle current $I^N$, while $C_2$ characterizes its zero-frequency fluctuations. 
Following Ref.~\cite{Ptaszynski2022}, the higher-order statistics, such as the skewness and kurtosis, are conveniently characterized through the dimensionless cumulant ratios
\begin{equation}
S\equiv\frac{C_3}{C_1},
\qquad
K\equiv\frac{C_4}{C_2}.
\label{eq:SK}
\end{equation}
The complete probability distribution can be recovered from the characteristic function through the inverse Fourier transform,
\begin{equation}
P_{\Delta t}(N)
=
\frac{1}{2\pi}
\int_{-\pi}^{\pi}
d\chi\,
\Lambda_{\Delta t}(\chi)e^{-i\chi N}.
\label{eq:PN}
\end{equation}

For a coherent two-terminal conductor of noninteracting electrons, whose average particle current is given by the Landauer formula~\cite{Blanter2000}, 
\begin{equation}
\label{eq:LB}
I^N = \frac{g_s}{h}\int \!d\varepsilon\, \mathrm{tr}\!\left[t^\dagger(\varepsilon)\,t(\varepsilon)\right]\left[f_L(\varepsilon) - f_R(\varepsilon)\right],
\end{equation}
where $t(\varepsilon)$ is the system energy-dependent transmission matrix, the trace is taken over the open channels in both the $R$ and $L$ contacts (see Fig.~\ref{fig:setup}), and $g_s=2$ accounts for spin degeneracy, the CGF is given by the Levitov--Lesovik formula~\cite{Levitov1993,Levitov1996}
\begin{align}
\mathcal{F}_{\Delta t}(\chi)
={}&
\frac{g_s\Delta t}{h}
\int d\varepsilon
\sum_p
\ln\Bigl\{
1
\nonumber\\
&+
\tau_p(\varepsilon)
\bigl(e^{i\chi}-1\bigr)
f_L(\varepsilon)
\bigl[1-f_R(\varepsilon)\bigr]
\nonumber\\
&+
\tau_p(\varepsilon)
\bigl(e^{-i\chi}-1\bigr)
f_R(\varepsilon)
\bigl[1-f_L(\varepsilon)\bigr]
\Bigr\},
\label{eq:lnlambda}
\end{align}
where 
%$g_s=2$ accounts for spin degeneracy, 
$p$ labels the transport eigenchannels and $\tau_p(\varepsilon)$ denote the corresponding energy-dependent transmission eigenvalues of the matrix $t^\dagger(\varepsilon)t(\varepsilon)$.
It is standard to call ${\cal T}(\varepsilon) \equiv \mbox{tr}[t^\dagger(\varepsilon)t(\varepsilon)] = \sum_p \tau_p(\varepsilon)$ the total transmission.
The Fermi distribution of reservoir $\alpha=L$ or $R$ is
\begin{equation}
f_\alpha(\varepsilon)
=
\frac{1}
{\exp\!\left[(\varepsilon-\mu_\alpha)/(k_BT_\alpha)\right]+1},
\end{equation}
where $\mu_\alpha$ and $T_\alpha$ are the chemical potential and
temperature, respectively. 
The terms proportional to $e^{i\chi}$ and $e^{-i\chi}$ account for forward ($L\to R$) and backward ($R\to L$) particle transfer events, respectively.
Equation~(\ref{eq:lnlambda}) therefore generates all cumulants of the transferred particle number and, through Eq.~(\ref{eq:PN}), determines its complete probability distribution.

Having reviewed the full counting statistics for particle transport, we now return to the statistics of entropy production. For simplicity, we specialize to equal reservoir temperatures, $T_L=T_R\equiv T$, and define the chemical potential bias as $\Delta\mu=\mu_L-\mu_R$. 
Under these conditions, the Fermi functions satisfy the local detailed-balance relation~\cite{PhysRevB.75.155316,Esposito2009,PhysRevB.78.115429}.
As a consequence, the CGF obeys the fluctuation symmetry $\mathcal{F}_{\Delta t}(\chi) = \mathcal{F}_{\Delta t} \left( -\chi+i\frac{\Delta\mu}{k_BT}\right)$, which leads to the exchange fluctuation theorem (EFT) for particle transport~\cite{Esposito2009,PhysRevB.75.155316,Andrieux_2009,PhysRevB.78.115429,RevModPhys.83.771},
\begin{equation}
\frac{P_{\Delta t}(N)}
     {P_{\Delta t}(-N)}
=
\exp\!\left(
\frac{\Delta\mu}{k_BT}N
\right).
\label{eq:particle_FT}
\end{equation}
Thus, particle transfer events against the chemical potential bias remain possible at finite temperature 
due to thermal fluctuations, but are exponentially suppressed relative to forward transfer events. 

Furthermore, the connection between particle transfer and entropy
production follows directly in the isothermal limit. The steady-state
average entropy production rate associated with particle transport is
~\cite{Seifert2012,Landi2021}
\begin{equation}
\dot{\Sigma}
=
\frac{\Delta\mu}{T}I^N,
\end{equation}
where the average particle current, $I^N$, is given by
Eq.~\eqref{eq:C1}. 
For a realization with a net particle transfer $N$ during 
%the 
a measurement interval $\Delta t$, the corresponding stochastic entropy production is
\begin{equation}
\Sigma_{\Delta t}(N)
=
\frac{\Delta\mu}{T}N.
\label{eq:entropy_transferred_particles}
\end{equation}
and its ensemble average satisfies
$\langle\Sigma_{\Delta t}\rangle=\dot{\Sigma}\Delta t$. The corresponding dimensionless stochastic entropy production is~\cite{PhysRevB.75.155316,Esposito2009}
\begin{equation}
\sigma_N
\equiv
\frac{\Sigma_{\Delta t}(N)}{k_B}
=
\mathcal{A}N,
\qquad
\mathcal{A}
\equiv
\frac{\Delta\mu}{k_BT},
\label{eq:entropy_affinity}
\end{equation}
where $\mathcal{A}$ is the dimensionless thermodynamic affinity.
Since $N$ is integer valued, $\sigma_N$ is also discrete, and the
one-to-one mapping $\sigma_N=\mathcal{A}N$ implies
\begin{equation}
P_{\Delta t}^{(s)}(\sigma_N)
=
P_{\Delta t}(N).
\label{eq:Ps_PN}
\end{equation}
The EFT can therefore be written directly in terms of entropy
production as~\cite{PhysRevB.75.155316,Andrieux_2009}
\begin{equation}
\frac{P_{\Delta t}^{(s)}(\sigma_N)}
     {P_{\Delta t}^{(s)}(-\sigma_N)}
=
e^{\sigma_N}.
\label{eq:FT_entropy_distribution}
\end{equation}
Negative particle transfer events, $N<0$, thus correspond to negative
entropy production. More generally, the EFT constrains the full
probability distribution, rather than only its low-order moments.

Equation~(\ref{eq:FT_entropy_distribution}) quantifies the asymmetry between positive and negative entropy production events. 
In the present isothermal two-terminal setup, entropy production is directly proportional to the net particle transfer $N$. 
At each energy, the forward and backward transfer processes are weighted by \(\tau(\varepsilon)f_L(\varepsilon)[1-f_R(\varepsilon)]\) and \(\tau(\varepsilon)f_R(\varepsilon)[1-f_L(\varepsilon)]\), respectively. 
Their ratio is therefore independent of the transmission probability and, by local detailed balance, is fixed solely by the thermodynamic affinity. 
Consequently, the fluctuation symmetry does not depend on the specific energy profile of \(\tau(\varepsilon)\). 
The coupling regime, by contrast, determines the transmission spectrum and hence the explicit form of both the CGF and the entropy production distribution. 
The coupling strength can therefore substantially modify the shape and moments of the entropy production statistics while strictly preserving the underlying fluctuation symmetry.

In what follows, we make this distinction explicit by deriving analytical expressions for the CGF and the associated probability distributions in the weak- and strong-coupling regimes, showing how the coupling-dependent transmission modifies the current and entropy production statistics while preserving the fluctuation symmetry imposed by the thermodynamic affinity.

%%%%%%%%%%%%%%%%%%%%%%%%%%%%%%%%%%%%%%%%%%%%%%%%%%%%%%%%%%%%%%%%%%%%%%%%%%
\section{Results and Discussion}
\label{sec:results}

\subsection{Weak-coupling regime}
\label{sec:results_wc}
We first consider the weak-coupling limit, in which $\Gamma=\Gamma_L+\Gamma_R\ll
% k_BT
\Delta$, 
where $\Gamma_L$ and $\Gamma_R$ denote the coupling strengths of the conductor to the left and right reservoirs, respectively, and $\Delta$ is the mean level spacing. Weak-coupling in mesoscopic transport is typically realized in the Coulomb blockade regime of quantum dots~\cite{Beenakker1991, Aleiner2002, Nazarov2009}, provided $\Gamma \ll U$, where $U$ is the system charging energy.
Under the condition $\Gamma \ll k_BT \ll \Delta$ \cite{Torres2003}, which is met in most Coulomb blockade experiments, transport occurs through a single, well-isolated resonant level, as illustrated in Fig.~\ref{fig:setup}(a). 
In this limit, one can calculate $I^N$ using Eq.~\eqref{eq:LB} \cite{Torres2003}.
The corresponding transmission probability takes the Breit--Wigner shape~\cite{Beenakker1991,Nazarov2009}
%\cite{PhysRevLett.54.1196,5390027,BEENAKKER19911}
%minitum between \Delta\mu e \T, referencia junction
\begin{equation}
\tau(\varepsilon)
=
\frac{\Gamma_L\Gamma_R}
{\left(\varepsilon-\varepsilon_0\right)^2
+\Gamma^2/4},
\label{eq:BW}
\end{equation}
where $\varepsilon_0$ is the resonant-level energy. 
Since the transmission varies over an energy scale set by $\Gamma$, whereas the Fermi functions vary over the thermal scale $k_BT$, the condition $\Gamma\ll k_BT$ allows the reservoir occupations to be approximated
as constant across the resonance, $f_\alpha(\varepsilon)\simeq f_\alpha(\varepsilon_0)$~\cite{Beenakker1991,Bagrets2003}.
Under this approximation, the energy integral in the CGF, Eq.~(\ref{eq:lnlambda}), can be evaluated analytically.

Substituting Eq.~(\ref{eq:BW}) into Eq.~(\ref{eq:lnlambda}), and evaluating the Fermi functions at the resonant energy yields
\begin{align}
\ln\Lambda(\chi)
={}&
\frac{g_{s}\Delta t}{h}
\int_{-\infty}^{\infty}\!\!d\varepsilon\,
\ln\!\Bigg[
1+
\frac{\Gamma_L\Gamma_R\,\bar A(\chi)}
{\left(\varepsilon-\varepsilon_0\right)^2+\Gamma^2/4}
\Bigg],
\label{eq:CGFweakIntermediate}
\end{align}
where
\begin{align}
\bar A(\chi)
={}&
\left(e^{i\chi}-1\right)
f_L(\varepsilon_0)
\left[1-f_R(\varepsilon_0)\right]
\nonumber\\
&+
\left(e^{-i\chi}-1\right)
f_R(\varepsilon_0)
\left[1-f_L(\varepsilon_0)\right].
\label{eq:AbarWeak}
\end{align}
For convenience, we introduce
\begin{equation}
\bar D(\chi)
=
\Gamma^2
+
4\Gamma_L\Gamma_R\,\bar A(\chi),
\label{eq:DbarWeak}
\end{equation}
so that the argument of the logarithm can equivalently be written as
\begin{align}
1+
\frac{\Gamma_L\Gamma_R\,\bar A(\chi)}
{\left(\varepsilon-\varepsilon_0\right)^2+\Gamma^2/4}
=
\frac{
\left(\varepsilon-\varepsilon_0\right)^2+\bar D(\chi)/4
}{
\left(\varepsilon-\varepsilon_0\right)^2+\Gamma^2/4
}.
\label{eq:logratioWeak}
\end{align}
This form makes the counting-field dependence of the CGF explicit and allows the subsequent differentiation to be carried out directly,
\begin{align}
\mathcal{J}(\chi)
\equiv
\frac{\partial\ln\Lambda(\chi)}
{\partial(i\chi)}
=
\frac{g_{s}\Delta t}{h}
\int_{-\infty}^{\infty}
d\varepsilon\,
\frac{\bar F(\chi)}
{\left(\varepsilon-\varepsilon_0\right)^2
+\bar D(\chi)/4},
\label{eq:NchiWeak}
\end{align}
where
\begin{align}
\bar F(\chi)
={}&
\Gamma_L\Gamma_R
\Big[
e^{i\chi}
f_L(\varepsilon_0)
\left(1-f_R(\varepsilon_0)\right)
\nonumber\\
&\qquad
-
e^{-i\chi}
f_R(\varepsilon_0)
\left(1-f_L(\varepsilon_0)\right)
\Big].
\label{eq:FbarWeak}
\end{align}
The functions $\bar F(\chi)$ and $\bar D(\chi)$ are not independent. From Eqs.~(\ref{eq:DbarWeak}) and~(\ref{eq:FbarWeak}), one directly
finds
\begin{equation}
\frac{\partial\bar D(\chi)}
{\partial(i\chi)}
=
4\bar F(\chi).
\label{eq:DderivativeWeak}
\end{equation}
The remaining energy integral is of the standard Lorentzian form,
\begin{equation}
\int_{-\infty}^{\infty}
\frac{d\varepsilon}
{\left(\varepsilon-\varepsilon_0\right)^2+\bar D(\chi)/4}
=
\frac{2\pi}{\sqrt{\bar D(\chi)}},
\label{eq:LorentzianWeak}
\end{equation}
which results in
\begin{equation}
\mathcal{J}(\chi)
=
\frac{g_{s}\Delta t}{h}
\,2\pi\,
\frac{\bar F(\chi)}
{\sqrt{\bar D(\chi)}}.
\label{eq:NchiWeakFinal}
\end{equation}

To recover the CGF, we use Eq.~(\ref{eq:DderivativeWeak}) to write
\begin{equation}
\frac{\bar F(\chi)}
{\sqrt{\bar D(\chi)}}
=
\frac{1}{2}
\frac{\partial\sqrt{\bar D(\chi)}}
{\partial(i\chi)}.
\label{eq:FoverDWeak}
\end{equation}
Thus, for spin-degenerate transport, 
%$g_{s}=2$, \Iv{therefore}, Eq.~(\ref{eq:NchiWeakFinal}) becomes
we write
\begin{equation}
\frac{\partial\ln\Lambda(\chi)}
{\partial(i\chi)}
=
\frac{\Delta t}{\hbar}
\frac{\partial\sqrt{\bar D(\chi)}}
{\partial(i\chi)}.
\label{eq:CGFderivativeWeak}
\end{equation}
Subsequent integration with respect to $i\chi$ yields
\begin{equation}
\ln\Lambda(\chi)
=
\frac{\Delta t}{\hbar}
\sqrt{\bar D(\chi)}
+
C.
\label{eq:CGFweakBeforeNormalization}
\end{equation}
The integration constant C is fixed by the normalization condition $\Lambda(0)=1$. 
Since $\ln\Lambda(0)=0$, we find $C=-(\Delta t/\hbar)\Gamma$. 
The resulting CGF in the weak-coupling regime reads
\begin{equation}
\ln\Lambda(\chi)
=
-
\frac{\Delta t}{\hbar}
\left[
\Gamma_L+\Gamma_R
-
\sqrt{\bar D(\chi)}
\right].
\label{eq:CGFweak}
\end{equation}
As expected, this expression is consistent with the full counting statistics of the resonant-level model in the weak-coupling regime previously obtained via master-equation approaches~\cite{Bagrets2003}. 
Here, 
% its derivation directly
the direct derivation of the CGF
from the Levitov--Lesovik scattering formulation provides the starting point for our analysis of the corresponding entropy production statistics.

Importantly, using the local detailed-balance relation~\cite{Esposito2009,PhysRevB.75.155316,PhysRevB.78.115429},
\begin{equation}
\frac{
f_L(\varepsilon_0)\left[1-f_R(\varepsilon_0)\right]
}{
f_R(\varepsilon_0)\left[1-f_L(\varepsilon_0)\right]
}
=
e^{\mathcal A},
\end{equation}
one readily finds that $\bar D(-\chi+i\mathcal A)=\bar D(\chi)$. Consequently, the CGF satisfies $\ln\Lambda(-\chi+i\mathcal A)=\ln\Lambda(\chi)$, explicitly demonstrating that the fluctuation symmetry is preserved in the weak-coupling regime.
Since all current cumulants are obtained from successive derivatives of $\ln\Lambda(\chi)$ with respect to the counting field, it completely determines the nonequilibrium statistics of particle transport. The corresponding particle transfer probability distribution is obtained by substituting Eq.~(\ref{eq:CGFweak}) into Eq.~(\ref{eq:PN}). 
Finally, using the mapping established in Eq.~(\ref{eq:entropy_affinity}), the same distribution directly determines the corresponding entropy production statistics.
\begin{figure}
\centering
\includegraphics[width=\linewidth]{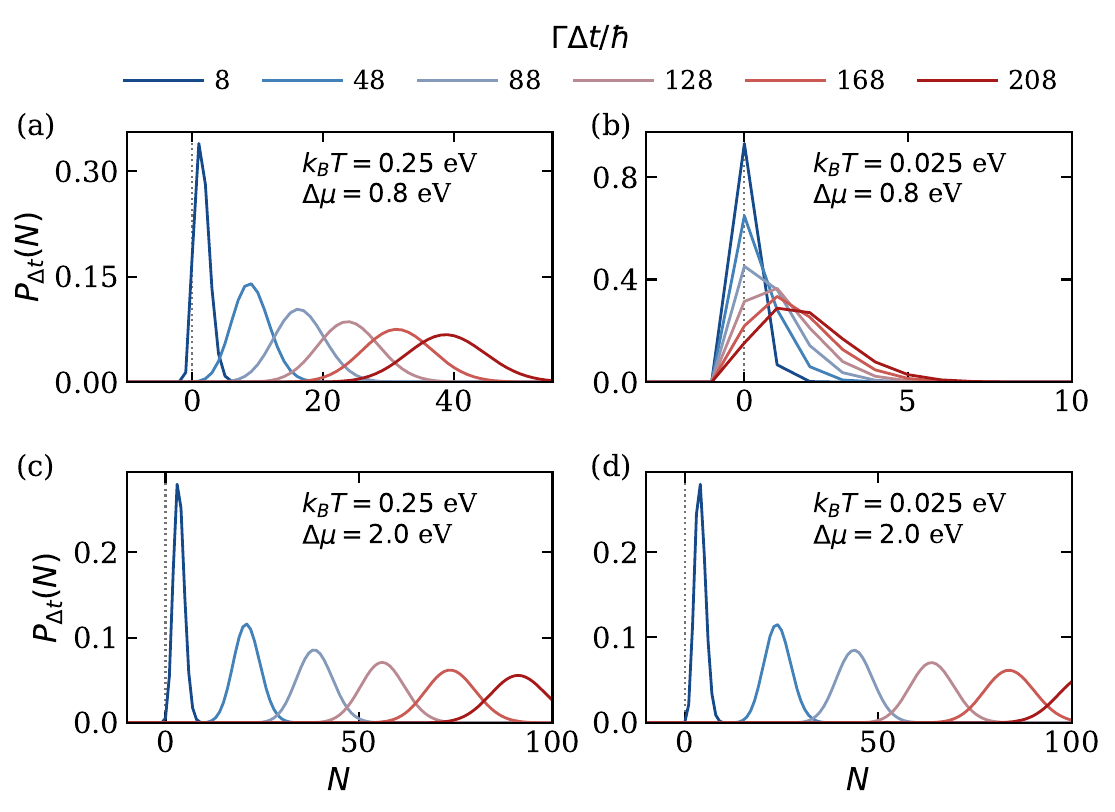}
\caption{
Probability distributions $P_{\Delta t}(N)$ of the net transferred particle number in the weak-coupling regime, with $\Gamma_L=\Gamma_R=0.00125$ eV ($\Gamma=0.0025$ eV).
Panels (a,b) correspond to a chemical potential difference $\Delta\mu=0.8$ eV and panels (c,d) correspond to $\Delta\mu=2.0$ eV.
The left column [(a,c)] shows results for $k_BT=0.25$ eV, while the right column [(b,d)] corresponds to $k_BT=0.025$ eV.
Different curves correspond to increasing dimensionless observation times $\Gamma\Delta t/\hbar$.
The vertical dotted line marks $N=0$, separating forward ($N>0$) from backward ($N<0$) particle transfer events.}
\label{fig:distribN_wc}
\end{figure}

To gain further insight into the weak-coupling CGF, we consider a single resonant level with $\varepsilon_0=-0.5$~eV and symmetric couplings $\Gamma_L=\Gamma_R=0.00125$~eV ($\Gamma=0.0025$~eV). 
We adopt a symmetric voltage drop around the reference chemical potential $\mu_0=0$, such that
$\mu_{L/R}=\mu_0\pm\Delta\mu/2$.
Figure~\ref{fig:distribN_wc} shows the resulting particle transfer distributions for different observation times, expressed in terms of the dimensionless quantity $\Gamma\Delta t/\hbar$.
Their behavior is governed by the interplay between the position of the resonant level relative to the transport window and the electronic occupations of the two reservoirs.
In the weak-coupling regime, the forward and backward transfer processes are weighted by $f_L(\varepsilon_0)[1-f_R(\varepsilon_0)]$ and $f_R(\varepsilon_0)[1-f_L(\varepsilon_0)]$, respectively.
These occupation factors control the activity of the corresponding transfer processes, whereas their ratio is fixed by the thermodynamic affinity through local detailed balance.

For $\Delta\mu=0.8$~eV [Figs.~\ref{fig:distribN_wc}(a,b)], the resonant level lies $0.1$~eV below $\mu_R$, i.e., nominally outside the bias window.
At $k_BT=0.25$~eV [Fig.~\ref{fig:distribN_wc}(a)], however, this detuning corresponds to only $0.4\,k_BT$, and the thermally broadened reservoir occupations remain appreciable at the resonant
energy~\cite{Datta1995}. 
Consequently, $f_L(\varepsilon_0)\simeq0.97$ and $f_R(\varepsilon_0)\simeq0.60$, so that the forward-transfer weight $f_L(\varepsilon_0)[1-f_R(\varepsilon_0)]\simeq0.39$ remains non-negligible, while the backward-transfer weight $f_R(\varepsilon_0)[1-f_L(\varepsilon_0)]\simeq0.02$ is strongly suppressed.
Thus, although the resonant level lies outside the nominal bias window, thermal broadening of the reservoir occupations allows it to contribute appreciably to particle transport.

At a lower temperature, [$k_BT=0.025$ eV, as shown in Fig.~\ref{fig:distribN_wc}(b)], the resonant level lies $4\,k_BT$ below $\mu_R$, and both reservoirs are nearly fully occupied at $\varepsilon=\varepsilon_0$, with $f_L(\varepsilon_0)\simeq 1$ and $f_R(\varepsilon_0)\simeq0.98$.
The forward-transfer factor $f_L(\varepsilon_0)[1-f_R(\varepsilon_0)]$ is small due to Pauli blocking in the right reservoir~\cite{Blanter2000}, and the distributions remain concentrated near $N=0$ as the observation time increases. Remarkably, this suppression occurs despite the larger thermodynamic affinity, $\mathcal{A}=32$, illustrating that transport activity and forward--backward asymmetry characterize distinct
aspects of the counting statistics.

For $\Delta\mu=2.0$ eV [Figs.~\ref{fig:distribN_wc}(c,d)], the resonant level lies inside the bias window.
Consequently, $f_L(\varepsilon_0)\simeq1$ and $f_R(\varepsilon_0)\ll1$, strongly favoring forward transfer through the resonance. 
The distributions therefore shift toward positive $N$ with increasing observation time.
In contrast to Figs.~\ref{fig:distribN_wc}(a,b), lowering the temperature does not suppress particle transfer, since the resonant level remains within the transport window. 
These results illustrate how the position of the resonant level relative to the bias window controls the transfer activity in the weak-coupling regime. 
\begin{figure}
    \centering
\includegraphics[width=\linewidth]{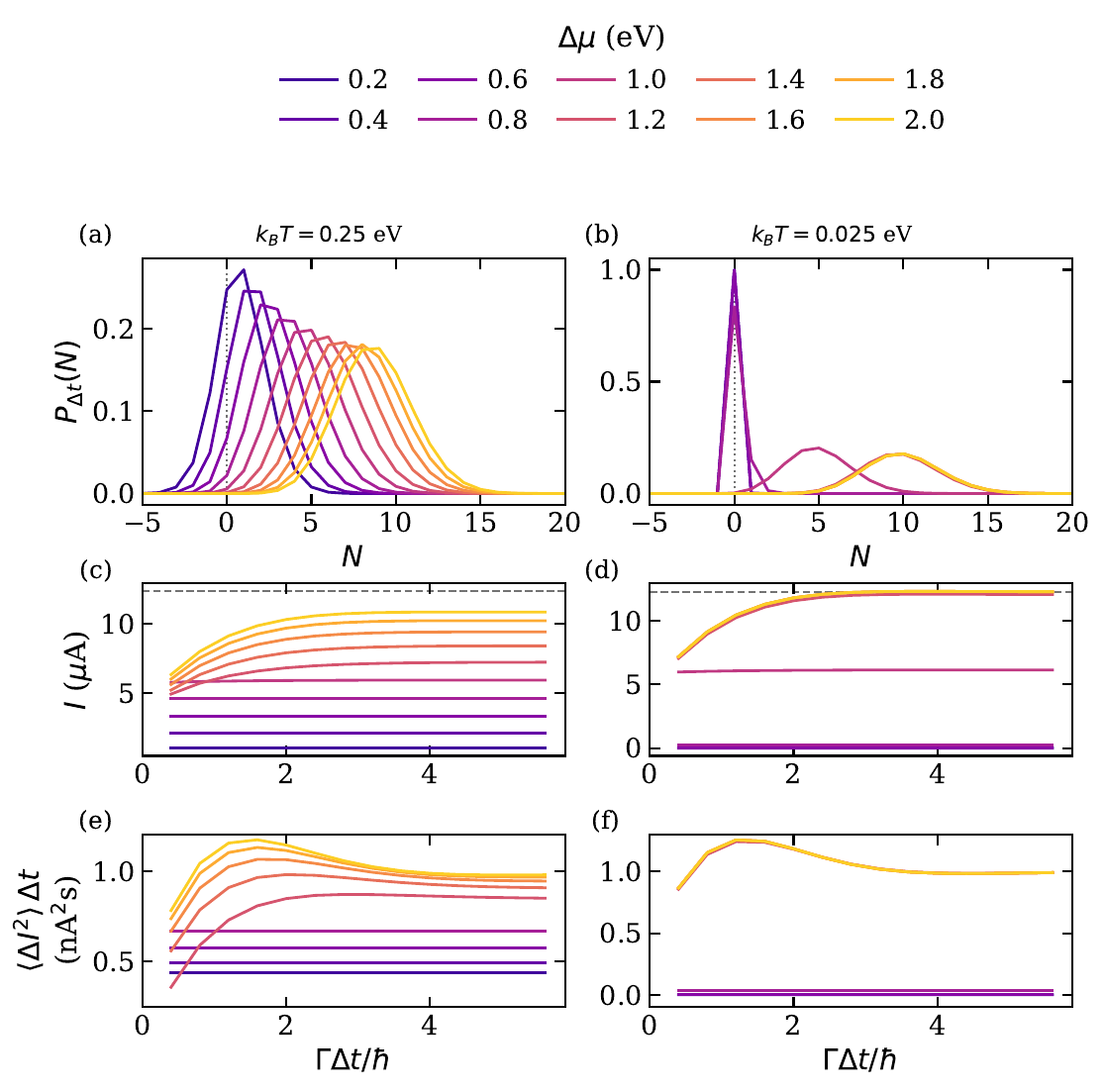}
\caption{Probability distributions $P_{\Delta t}(N)$ and the corresponding
current statistics in the weak-coupling regime, with
$\Gamma_L=\Gamma_R=0.00125$ eV
($\Gamma=0.0025$ eV).
Panels (a,b) show $P_{\Delta t}(N)$ for different source--drain
biases $\Delta\mu$ at a fixed dimensionless observation time
$\Gamma\Delta t/\hbar=20$.
Panels (c,d) show the average charge current, while panels (e,f)
show the current fluctuations as functions of the dimensionless
observation time $\Gamma\Delta t/\hbar$. 
Left-hand side panels show results for $k_BT = 0.25$~eV, while right-hand side panels [(b,d,f)] correspond to $k_BT = 0.025$~eV. 
Different curves correspond to increasing source--drain biases $\Delta\mu$, from $0.2$ to $2.0$ eV.}
\label{fig:distribDt_wc}
\end{figure}

Figure~\ref{fig:distribDt_wc} further illustrates the bias dependence of the weak-coupling counting statistics. 
At fixed $\Gamma\Delta t/\hbar=20$, the distributions shift smoothly toward positive $N$ with increasing $\Delta\mu$ at $k_BT=0.25$~eV [Fig.~\ref{fig:distribDt_wc}(a)], whereas at $k_BT=0.025$~eV
[Fig.~\ref{fig:distribDt_wc}(b)] they remain concentrated near $N=0$ until the resonant level enters the bias window at $\Delta\mu=2|\varepsilon_0|=1$~eV. 
The corresponding currents
[Figs.~\ref{fig:distribDt_wc}(c,d)] exhibit the same temperature
dependence, i.e., thermal broadening smooths the onset of resonant transport
at high temperature, whereas a sharp increase occurs at low temperature. At large
bias, the current approaches the weak-coupling resonant value
%$I\simeq(g_{s} e
$I^N = (g_s/\hbar)\Gamma_L\Gamma_R/(\Gamma_L+\Gamma_R)$~\cite{Blanter2000},
indicated by the dashed line in Fig.~\ref{fig:distribDt_wc}(d). The current fluctuations
[Figs.~\ref{fig:distribDt_wc}(e,f)] show a stronger bias dependence. They are suppressed below the resonant-transport threshold at low temperature,
while thermal broadening smooths this crossover at higher temperature.
\begin{figure}
    \centering
    \includegraphics[width=\linewidth]{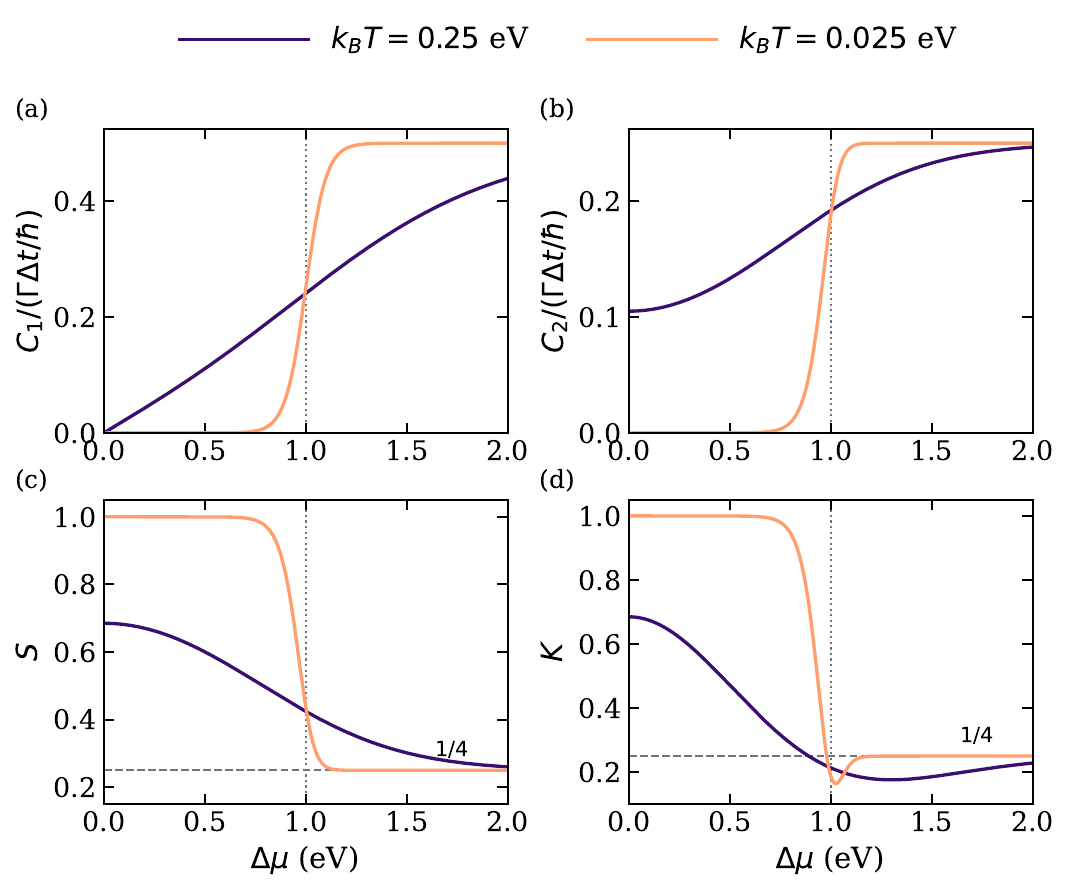}
    \caption{
Cumulants of the particle distribution in the weak-coupling regime as functions of bias voltage $\Delta\mu$. In all panels, $\Gamma_L = \Gamma_R = 0.00125$~eV ($\Gamma = 0.0025$~eV) and
$\varepsilon_0 = -0.5$~eV. 
(a) First cumulant (average particle current $I^N$) and (b) second cumulant (zero-frequency noise $S_2$) normalized per dimensionless observation time. 
%shows the first and (b) the second cumulants per dimensionless observation time. 
(c) 
% the normalized third and (d) fourth cumulants as functions of bias.
Dimensionless skewness $S$ (normalized third cumulant) and (d) dimensionless kurtosis $K$ (normalized fourth cumulant).
Purple curves correspond to $k_BT = 0.25$~eV and orange curves to $k_BT = 0.025$~eV. 
The vertical dotted line at $\Delta\mu = 1$~eV marks the threshold bias at which the resonant level enters the transport window for the symmetric voltage drop considered here. 
The horizontal dashed lines in panels (c,d) indicate the asymptotic large-bias limit
$S=K=1/4$ for symmetric coupling.}
    \label{fig:cumulants_wc}
\end{figure}

Figure~\ref{fig:cumulants_wc} shows the long-time cumulants in the weak-coupling regime. While $C_1$ and $C_2$ [Figs.~\ref{fig:cumulants_wc}(a,b)] reproduce the onset of resonant transport, the normalized higher-order cumulants $S=C_3/C_1$ and $K=C_4/C_2$~\cite{Ptaszynski2022} [Figs.~\ref{fig:cumulants_wc}(c,d)] characterize the associated change in counting statistics. 
To understand this behavior, we write
\begin{equation}
\bar{A}(\chi)
=
p_+\left(e^{i\chi}-1\right)
+
p_-\left(e^{-i\chi}-1\right),
\end{equation}
where
$p_+=f_L(\varepsilon_0)[1-f_R(\varepsilon_0)]$ and
$p_-=f_R(\varepsilon_0)[1-f_L(\varepsilon_0)]$.
Introducing $r=\Gamma_L\Gamma_R/\Gamma^2$, the weak-coupling CGF, Eq.(\ref{eq:CGFweak}), becomes
\begin{equation}
\ln\Lambda(\chi)
=
\frac{\Gamma\Delta t}{\hbar}
\left[
\sqrt{1+4r\bar{A}(\chi)}-1
\right].
\end{equation}
When transfer events are rare, $|4r\bar{A}(\chi)|\ll1$, expansion of
the square root gives
\begin{equation}
\ln\Lambda(\chi)
\simeq
\Delta t\left[
\gamma_+(e^{i\chi}-1)
+
\gamma_-(e^{-i\chi}-1)
\right],
\end{equation}
with $\gamma_\pm=(2\Gamma_L\Gamma_R/\hbar\Gamma)p_\pm$.
This is the CGF of two independent Poisson processes, whose difference follows a Skellam distribution~\cite{skellam}. 
At large affinity, $\gamma_-/\gamma_+=e^{-\mathcal A}\ll1$, backward transfers are suppressed and the statistics becomes effectively Poissonian, yielding $S,K\rightarrow1$. This accounts for the behavior below the resonant threshold in Fig.~\ref{fig:cumulants_wc}. 
Above the threshold, $p_+$ becomes of order unity and the rare-transfer expansion breaks down. The full CGF then yields, for symmetric coupling and large bias, $C_2/C_1\rightarrow 1/2$, $S\rightarrow 1/4$, $K\rightarrow 1/4$, consistent with sub-Poissonian resonant transport~\cite{Bagrets2003}.

%==================================================================
\subsection{Strong-coupling regime}
\label{sec:distribution_strong}
%==================================================================

In the strong-coupling regime, $\Gamma \agt \Delta$, the coupling-induced broadening can be sufficiently large that the transmission varies only weakly over the transport-relevant energy window set by $k_BT$ and $\Delta\mu$ \cite{Aleiner2002, Alves2002}. 
In this regime of strongly overlapping resonances, $\Gamma$ sets the characteristic energy scale for variations in ${\cal T}(\varepsilon)$, as illustrated in Fig.~\ref{fig:setup}(b). 
For open dots, where there are typical many open channels and $\Gamma \gg k_BT$, so that the channel transmission eigenvalues can be approximated as energy independent, $\tau_p(\varepsilon)\simeq\tau_p$~\cite{Blanter2000,Nazarov2009}.
In contrast to the weak-coupling regime, transport is no longer controlled by the position of a narrow resonant level relative to the bias window. 
Instead, the counting statistics is governed by the transmission probabilities $\tau_p$, which account for the coherent contributions of several resonances, and the thermodynamic affinity $\mathcal{A}$.

Under this approximation, the energy integration appearing in the Levitov--Lesovik CGF can be evaluated analytically~\cite{Nazarov2009}. Equation~(\ref{eq:lnlambda}) becomes
\begin{align}
\ln\Lambda(\chi)
={}&
\frac{g_{s}\Delta t}{h}
\int_0^\infty d\varepsilon
\sum_p 2
\ln\left[
\frac{\cosh(\beta\varepsilon)+A_p(\chi)}
{\cosh(\beta\varepsilon)+B}
\right],
\label{eq:CGFstrong1}
\end{align}
where $A_p(\chi)=\tau_p\cosh(\alpha+i\chi)+(1-\tau_p)\cosh\alpha$, $B=\cosh\alpha$, with $\alpha=\Delta\mu/2k_BT$. Introducing the dimensionless variable $x=\beta\varepsilon$, such
that $d\varepsilon=k_BT\,dx$, Eq.~(\ref{eq:CGFstrong1}) reduces to
\begin{align}
\ln\Lambda(\chi)
=
\Delta t\frac{g_{s} k_BT}{h}
\sum_p 2
\int_0^\infty dx\,
\ln\left[
\frac{\cosh x+A_p(\chi)}
{\cosh x+B}
\right].
\label{eq:CGFstrong2}
\end{align}
The remaining integral follows from the identity
\begin{equation}
\int_0^\infty dx\,
\ln\left[
\frac{\cosh x+\cosh a}
{\cosh x+\cosh b}
\right]
=
\frac{a^2-b^2}{2}.
\label{eq:logcosh_identity}
\end{equation}
Taking $a_p(\chi)=\operatorname{arccosh}A_p(\chi)$, $ b=\operatorname{arccosh}B=\alpha$, we obtain
\begin{align}
\ln\Lambda(\chi)
={}&
\Delta t\frac{g_{s} k_BT}{h}
\sum_p
\Biggl\{
\operatorname{arccosh}^{2}
\Biggl[
\tau_p
\cosh\left(
\frac{\Delta\mu}{2k_BT}+i\chi
\right)
\nonumber\\
&
+
(1-\tau_p)
\cosh\left(
\frac{\Delta\mu}{2k_BT}
\right)
\Biggr]
-
\left(
\frac{\Delta\mu}{2k_BT}
\right)^2
\Biggr\},
\label{eq:CGFstrong}
\end{align}
where the branch of $\operatorname{arccosh}$ is chosen by analytic continuation from $\chi=0$, ensuring $\ln\Lambda(0)=0$.
Within the energy-independent transmission approximation, Eq.~(\ref{eq:CGFstrong}) recovers the known finite-temperature full counting statistics of a coherent conductor with energy-independent transmission eigenvalues~\cite{Nazarov2009}.
Its dependence on the channel transparencies $\tau_p$ allows us to analyze how coherent partitioning shapes the particle transfer and entropy production statistics beyond the weak-coupling regime.

The fluctuation symmetry can also be verified explicitly from Eq.~(\ref{eq:CGFstrong}). 
Since $\mathcal{A}=\Delta\mu/(k_BT)=2\alpha$, under the transformation $\chi\rightarrow-\chi+i\mathcal{A}$ one finds
\begin{align}
A_p(-\chi+i\mathcal{A})
&=
\tau_p
\cosh\left[
\alpha+i(-\chi+i\mathcal{A})
\right]
+(1-\tau_p)\cosh\alpha
\nonumber\\
&=
\tau_p\cosh(-\alpha-i\chi)
+(1-\tau_p)\cosh\alpha
\nonumber\\
&=
A_p(\chi),
\label{eq:Ap_FT_strong}
\end{align}
where $\cosh(-z)=\cosh z$ has been used. Consequently,
\begin{equation}
\ln\Lambda(-\chi+i\mathcal{A})
=
\ln\Lambda(\chi),
\label{eq:FT_CGF_strong}
\end{equation}
which confirms that the closed-form CGF preserves the fluctuation symmetry within the strong-coupling approximation considered here.
%
%----------------------------- FIGURE 5 ---------------------
\begin{figure}
    \centering
    \includegraphics[width=\linewidth]
{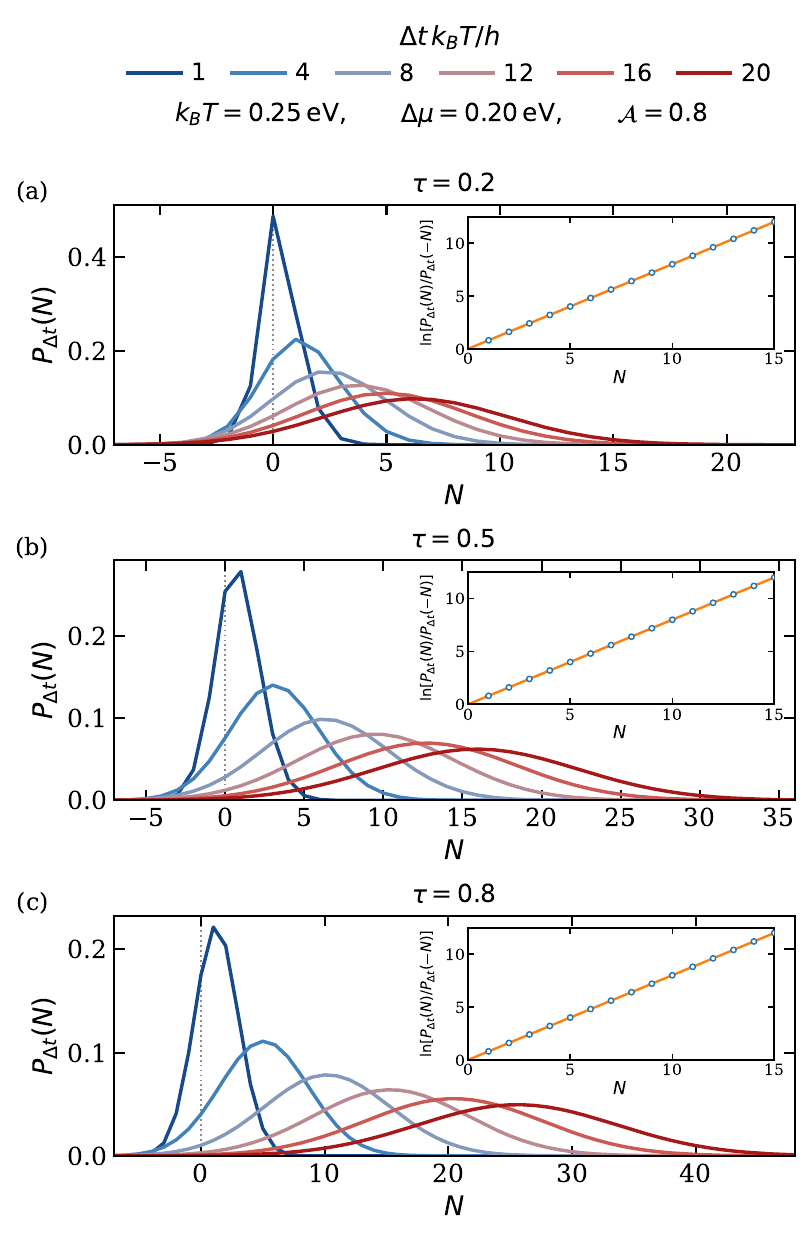}
    \caption{Probability distributions $P_{\Delta t}(N)$ of the net transferred particle number in the strong-coupling regime for transmission probabilities (a) $\tau=0.2$, (b) $\tau=0.5$, and (c) $\tau=0.8$.
    The distributions $P_{\Delta t}(N)$ are shown for different dimensionless observation times $\Delta t k_BT/h$, as indicated by the color scale. 
    The insets show $\ln[P_{\Delta t}(N)/P_{\Delta t}(-N)]$ (symbols) at $\Delta t k_BT/h=16$, together with the fluctuation-theorem prediction $\mathcal{A}N$ (solid line). 
    Here, $k_BT=0.25$ eV and $\Delta\mu=0.20$ eV ($\mathcal{A}=0.8$).}
    \label{fig:strongcoupling}
\end{figure}
%----------------------------- FIGURE 5 ---------------------

Figure~\ref{fig:strongcoupling} shows the resulting particle transfer distributions for a single transmission channel at fixed affinity $\mathcal{A}=0.8$ and three representative transparencies,
$\tau=0.2$, $0.5$, and $0.8$. 
As in the weak-coupling regime, the distributions progressively shift toward positive $N$ with increasing observation time, reflecting the positive mean particle current driven by the chemical potential
difference. At finite temperature, however, transfer events against the direction of the applied bias remain possible and give rise to the negative-$N$ tails of the distributions.

Increasing the transmission probability produces a progressively larger displacement of $P_{\Delta t}(N)$ toward $N>0$. 
The physical mechanism underlying this behavior differs from that in the weak-coupling regime. Here the transmission varies only weakly across the relevant energy interval, ensuring that the transport channels remain open throughout the transport window, where its transparency $\tau$ directly controls the transfer statistics.
% activity.

Despite these changes in the shape of the distribution,
the forward--backward asymmetry remains constrained by exactly the
same fluctuation relation. This is demonstrated in the insets of
Fig.~\ref{fig:strongcoupling}, where the numerical results satisfy
\begin{equation}
\ln\frac{P_{\Delta t}^{(s)}(\sigma_{_{N}})}
        {P_{\Delta t}^{(s)}(-\sigma_{_{N}})}
=
\sigma_{_{N}}
=
\mathcal{A}N
\label{eq:FT_entropy_strong}
\end{equation}
for all three transmission probabilities. Thus, changing the
transparency modifies the detailed counting statistics without
altering the fluctuation symmetry imposed by microscopic
reversibility. This result makes explicit the distinction already
identified in the weak-coupling regime: the microscopic transport
mechanism determines the shape and activity of the distribution,
whereas the thermodynamic affinity fixes the relative probability of
forward and backward entropy production fluctuations.

\begin{figure}[t]
    \centering
    \includegraphics[width=\linewidth]
    {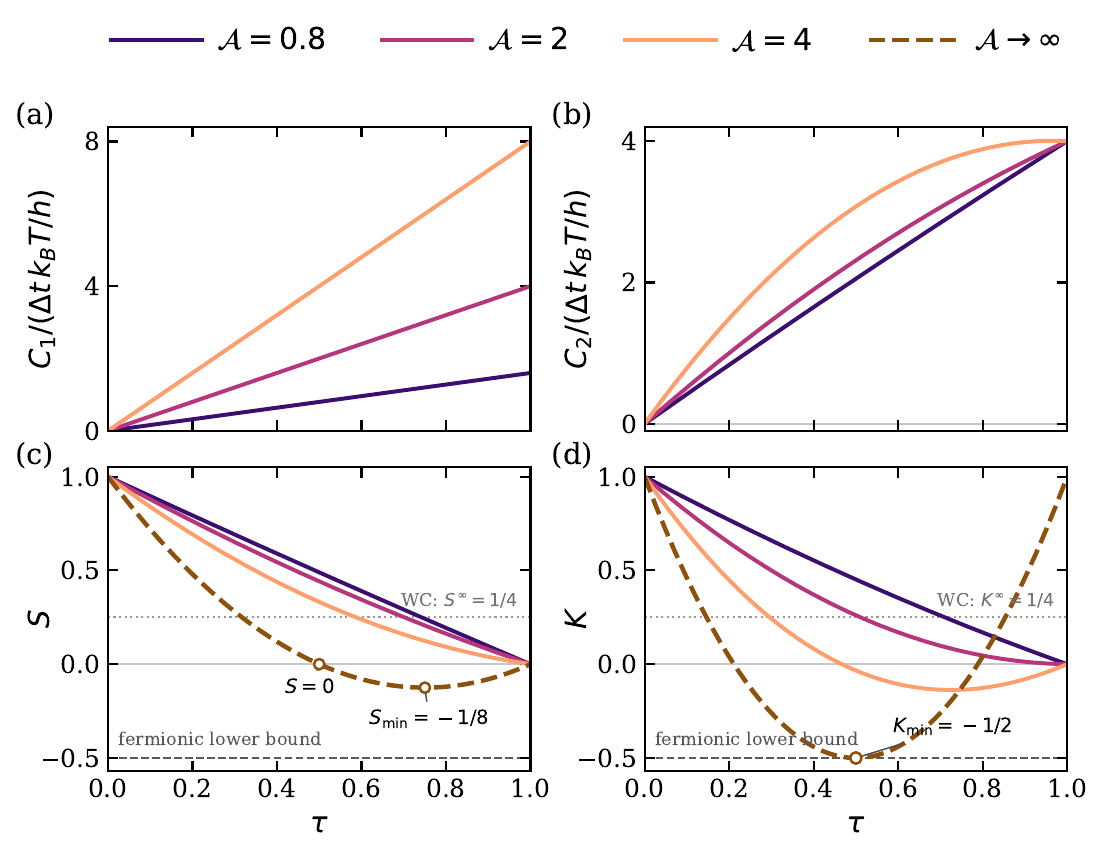}
    \caption{
    First four current cumulants in the strong-coupling regime for a single energy-independent transmission channel with transparency $\tau$.
    (a) First cumulant $C_1$ and (b) second cumulant $C_2$, as functions of the transmission probability $\tau$ for different values of the thermodynamic affinity ${\cal A}$.
    (c) Dimensionless skewness $S$ (normalized third cumulant) and (d) dimensionless kurtosis $K$ (normalized fourth cumulant). 
    The dashed red curves correspond to the large-bias limit $\mathcal{A}\rightarrow\infty$. 
    Horizontal dashed lines indicate the universal fermionic lower bound (-1/2), while the horizontal dotted lines mark the symmetric weak-coupling large-bias result $S_{\rm WC}^{\infty}=K_{\rm WC}^{\infty}=1/4$.}
    \label{fig:cumulants_sc}
\end{figure}

Successive derivatives of Eq.~(\ref{eq:CGFstrong}) with respect to $i\chi$ yield the cumulants. 
For a single channel, the first two are given by
\begin{align}
C_1
&=
\frac{g_{s}\Delta\mu}{h}\tau,
\label{eq:C1strong}
\\
C_2
&=
\frac{g_{s}}{h}
\left[
2k_BT\,\tau^2
+
\Delta\mu\,\tau(1-\tau)
\coth\left(
\frac{\Delta\mu}{2k_BT}
\right)
\right].
\label{eq:C2strong}
\end{align}
These expressions reproduce the standard finite-temperature current
and noise cumulants for an energy-independent transmission
channel~\cite{Blanter2000,Nazarov2009}, with $C_2$ containing both
thermal and partition-noise contributions.

The structure of the higher-order cumulants becomes particularly transparent in the large-affinity limit, $\mathcal{A}=\Delta\mu/(k_BT)\gg1$, which may be reached by taking $T\rightarrow0$ at fixed finite bias,
$\Delta\mu\neq0$. 
The latter condition preserves a finite transport window; if $\Delta\mu$ is simultaneously taken to zero, the transport cumulants vanish and the normalized ratios considered below become ill-defined.
In this limit, thermal contributions become subleading, and the first four cumulants reduce to
\begin{align}
C_1^{(\infty)}
&=
\frac{g_{s}\Delta\mu}{h}\tau,
\nonumber\\
C_2^{(\infty)}
&=
\frac{g_{s}\Delta\mu}{h}\tau(1-\tau),
\nonumber\\
C_3^{(\infty)}
&=
\frac{g_{s}\Delta\mu}{h}
\tau(1-\tau)(1-2\tau),
\nonumber\\
C_4^{(\infty)}
&=
\frac{g_{s}\Delta\mu}{h}
\tau(1-\tau)(1-6\tau+6\tau^2).
\label{eq:cumulants_strong_largeA}
\end{align}
These expressions highlight the role of fermionic partitioning: $C_{n\geq2}$ vanish as $\tau\rightarrow1$, while $C_3$ and $C_4$ exhibit sign changes at intermediate transmission. 
The corresponding normalized third and fourth cumulants are
\begin{equation}
\begin{aligned}
S^{(\infty)}
&\equiv
\frac{C_3^{(\infty)}}{C_1^{(\infty)}}
=
(1-\tau)(1-2\tau),
\\
K^{(\infty)}
&\equiv
\frac{C_4^{(\infty)}}{C_2^{(\infty)}}
=
1-6\tau+6\tau^2.
\end{aligned}
\label{eq:SKstrong}
\end{equation}
The above equations reveal a nonmonotonic dependence on the channel transparency. 
The third-cumulant ratio $S^{(\infty)}$ vanishes at $\tau=1/2$, becomes negative for $1/2<\tau<1$, and reaches its minimum, $S_{\min}^{(\infty)}=-1/8$, at $\tau=3/4$. Within the present single-channel large-affinity limit, this minimum is more restrictive than the general fermionic bound $S\geq-1/2$ derived in
Ref.~\cite{Ptaszynski2022}.

It is worth noticing that the above expressions also encompass the tunneling limit. For $\tau\ll 1$, the first four cumulants satisfy
\begin{equation}
C_n^{(\infty)}
\simeq
\frac{g_s\Delta\mu}{h}\tau,
\qquad n=1,\ldots,4,
\end{equation}
and hence
\begin{equation}
S^{(\infty)}\rightarrow1,
\qquad
K^{(\infty)}\rightarrow1.
\label{eq:Poisson_strong}
\end{equation}
In the large-affinity limit, backward transfers are exponentially suppressed, and transport through a single energy-independent channel reduces to a sequence of Bernoulli transmission attempts, with transmission probability $\tau$ and reflection probability $1-\tau$.
The resulting particle transfer statistics are therefore binomial~\cite{Levitov1993,Levitov1996,Nazarov2009}. 
In the tunneling limit, successful transmission events become rare and approximately independent, yielding the Poissonian limit in Eq.~(\ref{eq:Poisson_strong}). 
Away from this limit, fermionic partitioning gives rise to the non-Poissonian behavior of the higher-order cumulants.

Figure~\ref{fig:cumulants_sc} illustrates this behavior through the
first four cumulants obtained from Eq.~(\ref{eq:CGFstrong}). This
figure provides the broad-channel counterpart of
Fig.~\ref{fig:cumulants_wc}. Whereas in the weak-coupling regime the
bias controls whether the narrow resonance lies within the transport
window, here the channel transparency $\tau$ becomes the natural
parameter controlling the statistics.

Taken together, Figs.~\ref{fig:strongcoupling} and
\ref{fig:cumulants_sc} show that substantial changes in the transport
statistics can occur while the thermodynamic forward--backward
constraint remains unchanged. The coupling regime therefore controls
the detailed structure of irreversible fluctuations, while the
thermodynamic affinity controls their asymmetry.

%%%%%%%%%%%%%%%%%%%%%%%%%%%%%%%%%%%%%%%%%%%%%%%%%%%%%%
\section{Conclusions}
\label{sec:conclusions}
%%%%%%%%%%%%%%%%%%%%%%%%%%%%%%%%%%%%%%%%%%%%%%%%%%%%%%

We investigated the full statistics of entropy production in coherent mesoscopic transport beyond the weak-coupling limit. Within the Landauer--B\"uttiker framework, we related entropy production to the
full counting statistics of particle transfer and analyzed its probability distribution and higher-order cumulants. We considered two complementary transport regimes: a narrow resonant level in the weak-coupling limit and a broad channel with approximately energy-independent transmission in the strong-coupling limit. Although these regimes exhibit markedly different counting statistics, the fluctuation symmetry remains unchanged and is fixed solely by the thermodynamic affinity.

The two transport regimes, however, lead to distinct counting statistics. In the weak-coupling limit, the particle transfer statistics are controlled by the energetic accessibility of the narrow resonance, whereas in the strong-coupling limit they are governed by fermionic partitioning. This difference is particularly
visible in the higher-order cumulants, whose sign and nonmonotonic dependence on the transmission probability provide sensitive fingerprints of the underlying transport mechanism. Our results therefore show that the fluctuation theorem constrains the forward--backward asymmetry of entropy production independently of
these changes in the counting statistics, while higher-order cumulants retain information about the microscopic transport mechanism.

\acknowledgments
IRACL  acknowledges financial support from FAPESP (Grant No. 2025/25006-9). 
CHL acknowledges financial support from FAPERJ (Grant \#E-26/204.003/2024) and CNPq (Grant \#310717/2026-4).
ARR acknowledges financial support from FAPESP (Grants No.~2026/03311-7, No.~2025/25217-0, No.~025/24203-5, No.~2023/09820-2, and No.~2021/14375-0).

%%%%%%%%%%%%%%%%%%%%%%%%%%%%%%%%%%%%%%%%%%%%%%%%%%%%%%%%%%%%%%%

\bibliography{quantum_thermodynamics,mesoscopia}

\end{document}